\documentclass[aps, prd, amsmath, amssymb, floats, floatfix, superscriptaddress, nofootinbib, showpacs,twocolumn,notitlepage,10pt]{revtex4-1}
\usepackage{amssymb,amsmath}

\usepackage{graphicx}
\usepackage{amsmath}
\usepackage{amssymb}
\usepackage{amsfonts}
\usepackage{times}
\usepackage{xspace} 
\usepackage[usenames]{color}
\usepackage{dcolumn}
\usepackage{bm}
\usepackage{mathrsfs}
\usepackage{tikz}
\usepackage{appendix}
\usepackage{breqn}
\usepackage[normalem]{ulem}
\usepackage{epstopdf}

\usepackage[tracking=true]{microtype}
\SetTracking{}{500}
\SetTracking{encoding={*}, shape=sc}{40}

\makeatletter
\let\cat@comma@active\@empty
\makeatother

\begin{document}

\title{Self-force on a scalar charge in a circular orbit about a Reissner-Nordstr\"{o}m black hole}

\author{Jezreel Castillo}
\affiliation{National Institute of Physics, University of the Philippines, Diliman, Quezon City 1101, Philippines}
\affiliation{The Abdus Salam International Centre for Theoretical Physics, Trieste 34151, Italy}
\author{Ian Vega}
\affiliation{National Institute of Physics, University of the Philippines, Diliman, Quezon City 1101, Philippines}
\author{Barry Wardell}
\affiliation{School of Mathematics and Statistics,
University College Dublin, Belfield, Dublin 4, Ireland}

\begin{abstract}
Motivated by applications to the study of self-force effects in scalar-tensor theories of gravity, we calculate the self-force exerted on a scalar charge in a circular orbit about a Reissner-Nordstr\"{o}m black hole. We obtain the self-force via a mode-sum calculation, and find that our results differ from recent post-Newtonian calculations even in the slow-motion regime. We compute the radiative fluxes towards infinity and down the black hole, and verify that they are balanced by energy dissipated through the local self-force -- in contrast to the reported post-Newtonian results. The self-force and radiative fluxes depend solely on the black hole's charge-to-mass ratio, the controlling parameter of the Reissner-Nordstr\"{o}m geometry. They both monotonically decrease as the black hole approaches extremality. With respect to an extremality parameter $\epsilon$, the energy flux through the event horizon is found to scale as $\sim \epsilon^{5/4}$ as $\epsilon \rightarrow 0$.
\end{abstract}


\maketitle

\section{Introduction}    

The self-force acting on a particle moving in a curved spacetime has been a fascinating subject for some time, principally motivated by the prospect of detecting low-frequency gravitational waves with future space-based missions such as LISA \cite{amaro-seoane_2012}. While self-force-based gravitational waveforms remain elusive, progress in self-force research has been steady, and has made key contributions to a fuller understanding of the strongly-gravitating two-body problem \cite{poisson_2011,barack_2009,wardell_comp_2013}. Beyond direct applications to gravitational-wave astronomy, the self-force has proven useful as a theoretical probe of the nonlocal features of the spacetime in which the particle moves. More specifically, the static self-force has been shown to be sensitive to central and asymptotic structure \cite{drivas_2011,taylor_2013,kuchar_2013}. At the frontier of self-force research there remains strong momentum for the calculation of fully self-consistent gravitational waveforms from extreme-mass-ratio inspirals (to second order in the mass ratio), in addition to active research programs pushing to develop the formalism to higher dimensions \cite{taylor_2015,harte_2016,harte_2017} and alternative theories of gravity \cite{zimmerman_2015}. For the latter, it is of strong interest to understand how the extra gravitational degrees of freedom and their coupling might impact self-forced dynamics.

The work reported in this paper began with a consideration of self-forces in the context of alternative theories of gravity, particularly in scalar-tensor theories, as inspired by the seminal work of \citeauthor{zimmerman_2015} \cite{zimmerman_2015}. One tantalizing result of this work was the possibility of ``scalarization'' of a compact object by the action of the self-force. However, a close inspection of \cite{zimmerman_2015} quickly reveals that this effect requires a non-trivial scalar field residing on a curved spacetime background. No-hair theorems for black holes in scalar-tensor theories then severely limit the possible realizations for scalarization via self-force. To be sure, there are loopholes to these theorems; certain scalar-tensor theories do admit hairy black hole solutions \cite{torii_2001,kanti_1996,sotiriou_2014,antoniou_2018,antoniou_2017}. But these solutions are obtained mostly with numerical integration \cite{sotiriou_2014}, making them difficult to study as backgrounds for concrete self-force calculations. (See, however, \cite{babichev_2017_hairybhs} for some simple hairy black hole solutions in Horndeski theory.)

A well-known solution in scalar-tensor theory is the so-called Bocharova-Bronnikov-Melnikov-Bekenstein (BBMB) solution \cite{bocharova_1970,bekenstein_1974,bekenstein_1975} in conformal scalar-vacuum gravity. This theory is defined by the action 
\begin{equation}
	\mathcal{S} = \frac{1}{4\pi} \int d^4x \sqrt{-g} \left(\frac{R}{4} - \frac{1}{2}\nabla_\mu \Phi \nabla^\mu\Phi -\frac{1}{12}R\Phi^2\right),
\end{equation}
and the BBMB solution reads
\begin{equation}
	ds^2 = -\left(1-\frac{M}{r}\right)^2 dt^2 + \dfrac{dr^2}{\left(1-\dfrac{M}{r}\right)^2} + r^2 d\Omega^2
\end{equation}
with
\begin{equation}
	\Phi = \frac{\sqrt{3}M}{r-M}.
\end{equation}
The metric in this solution is the extremal Reissner-Nordstr\"{o}m solution, and the scalar field is non-trivial, though it clearly diverges at the putative event horizon at $r=M$. This divergence muddles the interpretation of $r=M$ as a true event horizon and, correspondingly, of the BBMB solution as a legitimate black hole solution. But this interpretational issue can be eschewed when one's primary concern is the impact of the scalar degree of freedom on the gravitational dynamics. This is the viewpoint espoused by our research program, of which this work is an initial step. Our proposal is to use the BBMB solution as a theoretical playground for studying scalar-tensor self-force effects. 

Apart from this broader goal, the Reissner-Nordstr\"{o}m spacetime is in itself an interesting spacetime on which to study self-force effects. As the unique, spherically-symmetric and asymptotically flat solution to the Einstein-Maxwell equations, it describes the spacetime outside a charged spherically-symmetric mass distribution. It is characterized by its mass $M$ and charge $Q$, and the spacetime is described by the metric
\begin{equation}
\label{eq:line-element}
ds^2 = -f(r) dt^2 + f(r)^{-1}dr^2 + r^2 d\theta^2 + r^2 \sin^2 \theta d\phi^2,
\end{equation}
where $f(r) = (r-r_+)(r-r_-)/r^2$ and $r_{\pm} = M \pm \sqrt{M^2 - Q^2}$. Note that the coordinate $r$ is connected to Q. If the object in question is a black hole, then it will have the following features:
\begin{enumerate}
  \item There are 2 horizons, an inner horizon at $r=r_-$, and an outer horizon at $r=r_+$, which happens to be the event horizon of the black hole.
  \item In the case where $M = Q$, the black hole becomes extremal, with a degenerate horizon (and thus zero temperature).
\end{enumerate}	
Despite these interesting properties, astrophysical considerations preclude significant charge build-up, and thus the self-force on a Reissner-N\"{o}rdstrom background has been largely neglected. The notable exception is a recent work by \citeauthor{bini_2016} \citep{bini_2016} in which they produced a 7 post-Newtonian (PN) order calculation of the scalar self-force on a circular geodesic. In this paper, we go beyond the PN approximation and present the first mode-sum calculation of the full strong-field scalar self-force on a circular geodesic. In the process of doing so, we found that our results for the self-force and energy flux is in disagreement with the slow-motion formulas presented by \citeauthor{bini_2016} \citep{bini_2016}. This is surprising, as one would expect a numerical calculation to agree with a PN calculation up to the order reported. We have yet to establish a reason for this discrepancy. Nevertheless, we present several consistency checks on our results, to show that this disagreement is not from an error in the numerical calculation.    

The paper is organized as follows: In Section II we provide a brief review of circular geodesics in Reissner-Nordstr\"{o}m spacetime. In Section III we provide a derivation of the scalar field generated by a scalar point charge in a geodesic circular orbit, subject to ingoing wave conditions at $r=r_+$ and outgoing wave conditions at $r=\infty$. In Section IV we briefly review the mode-sum regularization scheme, as well as providing a brief derivation of the regularization parameters for geodesic circular orbits in Reissner-Nordstr\"{o}m spacetime. In Section V we provide numerical results computed from a frequency-domain calculation and compare it with analytical results obtained from \citep{bini_2016}.

\section{Circular geodesics}

Considered as a test particle, the scalar charge will move along a geodesic of the Reissner-Nordstr\"{o}m spacetime. Two Killing vectors of spacetime, $t^\alpha := (\partial/\partial t)^\alpha$ and $\phi^\alpha := (\partial/\partial\phi)^\alpha$, provide the conserved quantities 
\begin{align}
	E &:= - g_{\alpha\beta}u^\alpha t^\beta = f(r) \frac{dt}{d\tau} \\
	L &:= g_{\alpha\beta}u^\alpha \phi^\beta = r^2 \sin^2\theta \frac{d\phi}{d\tau},
\end{align}
where $\tau$ is the proper time along the orbit and $u^\alpha$ is the particle's four-velocity. 

Due to the spherical symmetry of the spacetime, the test particle will move along a fixed plane. We can always choose our coordinates so that this plane is described by $\theta = \pi/2$. Combining these with the normalization, $u^\alpha u_\alpha =-1$, we arrive at the radial equation
\begin{equation}
	\left(\frac{dr}{d\tau}\right)^2 = E^2 - V(r), 
	\label{eqn:radialeqn}
\end{equation}
where $V(r) := f(r)(1+L^2/r^2)$ is the effective potential. For circular orbits ($r=r_0$), the four-velocity reads 
\begin{equation}
	u^\alpha = \frac{dt}{d\tau}(1,0,0,\Omega),
\end{equation}
where 
\begin{equation}
	\Omega := d\phi/dt = (L/E)f(r_0)/r_0^2
	\label{eqn:angvel}
\end{equation}
is the angular velocity of the particle with respect to an asymptotic observer. Note that this quantity, as an observable, is invariant to coordinate transformations. Circular orbits also require $V'(r_0) =0$, which gives the condition 
\begin{equation}
	L^2 = \frac{r_0^2\left(Mr_0-Q^2\right)}{r_0^2-3Mr_0+2Q^2}, 
\end{equation}
while $dr/d\tau =0$ in Eq.~(\ref{eqn:radialeqn}) gives 
\begin{equation}
	E^2 = \frac{\left(r_0^2-2Mr_0+Q^2\right)^2}{r_0^2\left(r_0^2-3Mr_0+2Q^2\right)},
\end{equation}
which can be combined to give 
\begin{equation}
	\left(\frac{L}{E}\right)^2 = \frac{r_0^4\left(Mr_0-Q^2\right)}{\left(r_0^2-2Mr_0+Q^2\right)^2}.
\end{equation}
Putting this into Eq.~(\ref{eqn:angvel}) we finally get 
\begin{equation}
	\Omega^2 = \frac{M}{r_0^3}\left(1-\frac{Q^2}{Mr_0}\right).
	\label{eqn:angfreq}
\end{equation}
Normalization of the four-velocity then gives
\begin{equation}
		(u^t)^2 = \frac{r_0^2}{\left(r_0^2-3Mr_0+2Q^2\right)}.
\end{equation}
This completes the determination of the four-velocity for a particle in a circular geodesic. 

\section{Field equations}

\subsection{Multipole decomposition}

We assume that the scalar field $\Phi$ is a small perturbation of the fixed Reissner-Nordstr\"{o}m spacetime, and that it satisfies the minimally coupled scalar wave equation
\begin{equation}
	\Box \Phi = - 4\pi \mu,
	\label{waveeqn}
\end{equation}
sourced by a scalar charge density $\mu$. We model this scalar charge density as a $\delta$-function distribution on the worldline, written as
\begin{equation}
	\mu(x) = q \int \frac{\delta^{(4)}(x^\alpha-z^\alpha(\tau))}{\sqrt{-g}} \,d\tau,
\end{equation}
which for a circular orbit becomes
\begin{align}
	\mu(x) &= \frac{q}{r_0^2} \int \delta(t-t(\tau))\delta(r-r_0)\delta(\theta-\pi/2)\delta(\phi-\Omega t(\tau)) \,d\tau \\
	       &= \frac{q}{r_0^2u^t} \delta(r-r_0)\delta(\theta-\pi/2)\delta(\phi -\Omega t).
\end{align}
Using the spherical harmonic completeness relations, we can further rewrite $\mu$ as
\begin{align}
      \mu(x) &= \frac{q}{r_0^2u^t}\sum_{l,m}Y^{*}_{lm}(\pi/2,\Omega t)Y_{lm}(\theta,\phi)\delta(r-r_0) \\
      		 &= \sum_{l,m}\left[\frac{qY^{*}_{lm}(\pi/2,\Omega t)}{r_0^2u^t}\right]Y_{lm}(\theta,\phi)\delta(r-r_0) \\
             &= \sum_{l,m}\left[\frac{qY^{*}_{lm}(\pi/2,0)}{r_0^2u^t}\right]e^{-im\Omega t}Y_{lm}(\theta,\phi)\delta(r-r_0). \label{eq:SCDec}
\end{align}
A similar decomposition for the scalar field $\Phi$ into spherical harmonics and Fourier modes yields the form
\begin{align}
	\Phi = \int\sum_{l,m}\Phi_{lm}(r) e^{i\omega t}Y_{lm}(\theta,\phi) \ d\omega. \label{eq:SFDec}
\end{align}
With Eqs.~\eqref{eq:SCDec} and \eqref{eq:SFDec}, Eq.~(\ref{waveeqn}) reduces to
\begin{align}
	\frac{1}{r^2}\frac{d}{dr}\left( r^2 f(r)\frac{d\Phi_{lm}}{dr}\right) & + \left(\frac{\omega^2}{f(r)}-\frac{l(l+1)}{r^2}\right) \Phi_{lm} \nonumber \\ &= -\frac{qY^{*}_{lm}(\pi/2,0)}{r_0^2u^t} \delta(r-r_0) 
	\label{eqn:ode}
\end{align}
We now want to impose boundary conditions. 

\subsection{Boundary conditions}

The wave equation can also be rewritten in terms of the so-called tortoise coordinate 
\begin{align}
r_* :=& \int f^{-1} dr \nonumber \\ =& r+\frac{r_{+}^2}{r_{+}-r_{-}}\ln\left(r-r_{+}\right)-\frac{r_{-}^2}{r_+ - r_-} \ln\left(r-r_-\right). 
\end{align}
Defining $\Phi(t,r,\theta^A) = \Psi(t,r,\theta^A)/r$ (where $\theta^A := (\theta,\phi)$) we get 
\begin{equation}
	-\frac{\partial^2\Psi}{\partial t^2} + \frac{\partial^2\Psi}{\partial r_*^2} + f\left(\frac{\nabla_\Omega^2}{r^2}-\frac{f'}{r}\right)\Psi = -4\pi r f(r)\mu(x) 
	\label{eqn:tortcoordseqn}
\end{equation}
where $\nabla_\Omega^2$ is the Laplacian on the unit two-sphere. Decomposing $\Psi$ into its spherical-harmonic components
\begin{equation}
	\Psi(t,r,\theta^A) = \sum_{lm} \Psi_{lm}(t,r) Y_{lm}(\theta^A),
\end{equation} 
Eq.~(\ref{eqn:tortcoordseqn}) becomes
\begin{align}
	-\frac{\partial^2\Psi_{lm}}{\partial t^2} &+ \frac{\partial^2\Psi_{lm}}{\partial r_*^2} - f\left(\frac{l(l+1)}{r^2}+\frac{f'}{r}\right)\Psi_{lm} \nonumber \\ &= -\frac{4\pi q f(r_0)}{r_0 u^t} Y^*_{lm}(\pi/2,0)e^{i\omega_m t} \delta(r-R).
	\label{eqn:odetort}
\end{align}
The homogeneous part of this equation appears like a flat-space wave equation [in (1+1) dimensions] with a potential $V_l(r):= -f (l(l+1)/r^2 + f'/r)$. This potential vanishes as $r\rightarrow M$ (or as $r_* \rightarrow -\infty$) and as $r\rightarrow \infty $ ($r_* \rightarrow +\infty$). 

The appropriate boundary conditions are ingoing waves at the event horizon and outgoing waves at infinity. Since $\Psi_{lm} \sim e^{i\omega t}$ (where $\omega = -m\Omega$ for circular orbits), we shall then impose that $\Psi_{lm} \sim e^{-i\omega r_*}$ as $r\rightarrow \infty$ and $\Psi_{lm} \sim e^{i\omega r_*}$ as $r\rightarrow r_+$. Correspondingly, for $\Phi_{lm}(r)$ the boundary conditions of interest are
\begin{equation}
	\Phi_{lm} \sim \frac{e^{i\omega r_*}}{r},\,\, r \rightarrow r_+ 
\end{equation} 
and 
\begin{equation}
	\Phi_{lm} \sim \frac{e^{-i\omega r_*}}{r},\,\, r \rightarrow \infty.
\end{equation} 

These boundary conditions serve as initial data in the integration of Eq.~(\ref{eqn:ode}). In practice, the integration cannot begin exactly at the horizon because $f(r)$ vanishes and the potential term in Eq.~(\ref{eqn:ode}) blows up. [The potential term of Eq.~(\ref{eqn:odetort}) is regular, but the horizon in these coordinates is inaccessible at $r_* =-\infty$.] Instead, we then begin the integration slightly away from the horizon, at $r = r_+ +\varepsilon$ for $\varepsilon/r_+  \ll 1$. An asymptotic solution as $r\rightarrow r_+$ can be obtained by inserting the ansatz
\begin{equation}
	\Phi_{lm}(r) = \frac{e^{i\omega r_*}}{r} \sum_{n=0} a_n(r-r_+)^n
\end{equation}
into Eq.~(\ref{eqn:ode}). This gives a recurrence relation for the coefficients $a_n$ which reads
\begin{widetext}
\begin{multline}
	a_n = -\frac{8i\omega(n-1)r_+^3 + (6(n-1)(n-2) - \lambda)r_+^2 - 2M(3n^2-11n+9)r_+ + Q^2(n-2)(n-3)}{2nr_+(i\omega r_+^3+2(n-1)r_+^2-M(3n-4)r_+ + Q^2(n-2))}a_{n-1} \\ - \frac{6i\omega(n-2)r_+^2+(2(n-2)(n-3) - \lambda)r_+ - M(n-3)^2}{nr_+(i\omega r_+^3+2(n-1)r_+^2-M(3n-4)r_+ + Q^2(n-2))}a_{n-2}\\ - \frac{8i\omega(n-3)r_+ + (n-3)(n-4) -\lambda}{2nr_+(i\omega r_+^3+2(n-1)r_+^2-M(3n-4)r_+ + Q^2(n-2))}a_{n-3} \\ - \frac{i\omega (n-4)}{nr_+(i\omega r_+^3+2(n-1)r_+^2-M(3n-4)r_+ + Q^2(n-2))}a_{n-4}
\end{multline}
\end{widetext}
The same considerations apply to the boundary condition as $r\rightarrow \infty$. Again we work with the ansatz 
\begin{equation}
	\Phi_{lm}(r) = \frac{e^{-i\omega r_*}}{r} \sum_{n=0} \frac{b_n}{r^n},
\end{equation}
and obtain a recurrence relation for $b_n$ using Eq.~(\ref{eqn:ode}). This reads
\begin{align}
	b_n =& -\frac{n(n-1) - \lambda}{2i\omega n}b_{n-1} + \frac{M(n-1)^2}{i\omega n}b_{n-2} \nonumber \\ & - \frac{Q^2(n-1)(n-2)}{2i\omega n}b_{n-3}.
\end{align}

\section{Regularization}

\subsection{Mode-sum regularization}

To obtain the scalar self-force, we must first subject the unregularized force to a regularization procedure. In our case, we use the mode-sum scheme \cite{barack_2000,burko_2000_b,barack_2003}, where the self-force is constructed from regularized spherical harmonic contributions. We start with full force derived from the retarded field
\begin{equation}
F_{\alpha}^{\mathrm{full}}(x) = q\nabla_{\alpha}\Phi(x) = \sum_{l}F_{\alpha}^{(\mathrm{full}),l}(x),
\end{equation}
where $F_{\alpha}^{(\mathrm{full}),l}(x)$ is the $l$-mode component (summed over $m$) of the full force at an arbitrary point $x$ in the neighborhood of the particle. At the particle location, each $F_{\alpha}^{(\mathrm{full}),l}$ is finite, although the sided limits often produce different values (which we then label as $F_{\alpha,\pm}^{(\mathrm{full}),l}$) and the sum over $l$ may not converge. We then obtain the self-force using a mode-by-mode regularization formula
\begin{multline}
F_{\alpha}^{\mathrm{self}} = \sum_{l} F_{\alpha}^{\mathrm{(full)},l} \\ = \sum_{l} (F_{\alpha,\pm}^{(\mathrm{full}),l}-A_{\alpha,\pm}(l+1/2)-B_{\alpha,\pm}),
\end{multline}
where the regularized contributions $F_{\alpha}^{(\mathrm{full}),l}$ no longer have the $\pm$ ambiguity and the sum over $l$ is guaranteed to converge. The regularization parameters ($l$-independent) $A_{\alpha}$ and $B_{\alpha}$ have been obtained for generic orbits about a Schwarzschild black hole \cite{barack_2002}, and a Kerr black hole \cite{barack_2003}. In the next subsection we present a derivation of $A_{\alpha}$ and $B_{\alpha}$ for circular orbits about a Reissner-Nordstr\"{o}m black hole.

\subsection{Regularization parameters}

The procedure for deriving mode-sum regularization parameters is by now well-established
\cite{barack_2002,barack_2003,Haas:2006ne,Heffernan:2012su,Heffernan:2012vj,Heffernan2014}. Here,	 we directly
follow the approach of \cite{Heffernan:2012su,Heffernan:2012vj,Heffernan2014}, extending it to the case of
Reissner-Nordstr\"{o}m spacetime as given by the line element, Eq.~\eqref{eq:line-element}. Since the
essential details remain the same, we refer the reader to
Refs.~\cite{Heffernan:2012su,Heffernan:2012vj,Heffernan2014} for an extensive discussion, and give here only the
key equations and results.

We start with an expansion of the Detweiler-Whiting singular field \citep{detweiler_2003} through next-from-leading order in the distance from the worldline,
\begin{align}
  \Phi^S \approx \frac{1}{\rho}
    - \frac{1}{\rho ^3} \bigg[&
      \frac{\Delta r \left[2 \Delta w_1^2 L^2+r_0^2 \left(\Delta w_1^2+\Delta w_2^2\right)\right]}{2 r_0} \nonumber \\ &
      - \frac{\Delta r^3 f'_0}{4 f_0^2}
      - \frac{E L (2 f_0 + r_0 f_0') \Delta t \Delta r \Delta w_1}{2 r_0 f_0} \nonumber \\ &
      + \frac{\Delta t^2 \Delta r \left(2 E^2-f_0\right) f_0'}{4 f_0}
    \bigg].
\end{align}
Here, we have already specialized to the case of circular, equatorial orbits, and have introduced $f_0 := f(r_0)$ and
\begin{align}
  \rho^2 := & \frac{\Delta r^2}{f_0}+\Delta w_1^2 \left(L^2+r_0^2\right)+\Delta w_2^2 r_0^2
  \nonumber \\
  & \quad
  + \left(E^2-f_0\right) \Delta t^2 -2 E L \Delta t \Delta w_1 .
\end{align}
The above expressions are given in terms of the Riemann normal coordinates $w_1$ and $w_2$, the same
as are described in \cite{Heffernan:2012vj}.

It turns out that for circular orbits the $t$ and $\phi$ componets of the self-force are purely dissipative, meaning that only the radial component of the self-force requires regularization. We thus
compute the contribution from the singular field to the radial component of the self-force using
$F_r^{\text S} = \partial_{\Delta r} \Phi^S$. Doing so, taking $\Delta t \to 0$, and keeping only
terms which will not vanish in the limit $\Delta r \to 0$ we get
\begin{equation}
  F_r^{\text S} = \frac{r_0}{2 L^2 \rho}\bigg(1-\frac{1}{\chi }\bigg)-\frac{\Delta r}{\rho ^3 f_0}
\end{equation}
Here, $\chi := 1 -  \tfrac{L^2}{r_0^2+L^2} \sin^2 \beta$, just as in \cite{Heffernan:2012vj}.

Next, we obtain the regularization parameters by decomposing this into spherical-harmonic modes (as
usual, we only need to consider the $m=0$ case since the other $m$-modes do not contribute). Doing
so, and taking the limit $\Delta r \to 0$, we find
\begin{equation}
  F_{r,\pm}^{S, l} = \mp \frac{2l+1}{2 r_0 f_0^{1/2} \sqrt{L^2 + r_0^2}} + \frac{\mathcal{E} - 2 \mathcal{K}}{\pi r_0 \sqrt{r_0^2 + L^2}},
\end{equation}
from which we can immediately read off the $A_r$ and $B_r$ regularization parameters. Here,
\begin{align}
  \mathcal{K} &:= \int_0^{\pi/2} (1 - \tfrac{L^2}{r_0^2+L^2} \sin^2 \beta)^{-1/2} d \beta, \nonumber \\
  \mathcal{E} &:= \int_0^{\pi/2} (1 - \tfrac{L^2}{r_0^2+L^2} \sin^2 \beta)^{1/2} d \beta
\end{align}
are complete elliptic integrals of the first and second kind, respectively.

\section{Self-force calculation}

\subsection{Scalar energy flux} 

Global energy conservation dictates that the local energy dissipation, represented by the $t$-component of the self-force, is accounted for by the energy flux carried by scalar field radiation. We numerically calculate the energy flux to infinity and down the black hole, and verify that the result is consistent with the energy lost through the local dissipative self-force.

We briefly review the relevant formalism used to calculate the energy flux. The stress-energy tensor of the scalar field is given by
\begin{equation}
	T_{\alpha \beta} = \frac{1}{4\pi}\left( \Phi_{;\alpha}\Phi_{;\beta}-\frac{1}{2}g_{\alpha \beta} \Phi^{;\mu}\Phi_{;\mu}\right).
\end{equation}
With $T_{\alpha \beta}$, we construct the differential energy flux over 
the following constant $r$ hypersurfaces: $r \to \infty$, represented by $\Sigma_+$, and $r \to r_+$, represented by $\Sigma_-$. The differential energy flux then takes the form
\begin{equation}
	d\mathcal{E}_{\pm} = \mp T_{\alpha \beta}t^{\alpha}n^{\beta}_{\pm}d\Sigma_{\pm}, \label{eq:DEF}
\end{equation}
where $n^{\alpha}$ is the unit normal vector of the hypersurface, and $d\Sigma$ is the hypersurface element. We then rewrite Eq.~\eqref{eq:DEF} as
\begin{equation}
	d\mathcal{E}_{\pm} = \mp T_{tr}f(r)r^2 dt d\Omega. \label{eq:diffEflux}
\end{equation}
Integrating over the two-sphere, we then express the energy transfer as
\begin{equation}
	\frac{dE_{\pm}}{dt}=\dot{E}_{\pm} = \mp \oint T_{tr}f(r)r^2d\Omega. \label{eq:Etransfer}
\end{equation}
Substituting the multipole expansion defined by Eq.~\eqref{eq:SFDec} into Eq.~\eqref{eq:Etransfer}, we then arrive at the following expression for the energy transfer
\begin{align}
\dot{E}_{\pm`'} = \pm i\frac{f(r)r^2}{4\pi}\sum_{l,m}\omega_{m}\Phi^{*}_{lm}\Phi_{lm,r}.
\end{align}

\begin{figure}[tb]
\includegraphics[width=\linewidth]{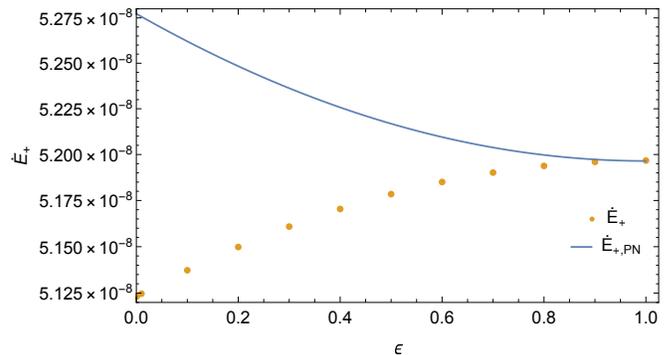}
\caption{Linear-log plot of $\dot{E}_{+}$ vs $\epsilon$ for $(M\Omega)^{-2/3}=50$. The orange dots represent the numerical results, while the solid blue line represents the results obtained from \citeauthor{bini_2016}'s slow motion formula for the outgoing flux. Disagreement between the two results increases as $\epsilon \to 0$.}
\label{fig:EFI}
\end{figure}

\begin{figure}[tb]
\includegraphics[width=\linewidth]{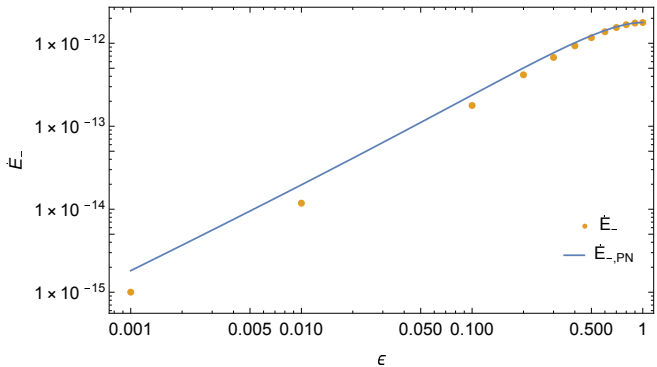}
\caption{Log-log plot of $\dot{E}_{-}$ vs $\epsilon$ for $(M\Omega)^{-2/3} = 50$. The orange dots represent the numerical results, while the solid blue line represents the results obtained from Bini \textit{et. al.}'s slow-motion formula for the horizon flux. For near-extremal values of $\epsilon$, $\dot{E}_{r=r_+}$ exhibits power law fall-off in $\epsilon$, which in this example is found to scale as  $\dot{E}_{r=r_+} \sim \epsilon^{5/4}$.}
\label{fig:EFH}
\end{figure}

	We present sample numerical data for $\dot{E}_{+}$, and $\dot{E}_{-}$ in Tables \ref{tab:EFI}, and \ref{tab:EFH} respectively. We see that as the extremality parameter $\epsilon:=1-Q/M$ approaches zero, $\dot{E}_{\pm}$ monotonically decreases. We also note that compared to $\dot{E}_{+}$, $\dot{E}_{-}$ exhibits a dramatic decrease as $\epsilon \to 0$. We then investigate the scaling behavior of $\dot{E}_{\pm}$ with respect to $\epsilon$, which we present in Fig. \ref{fig:EFI} and \ref{fig:EFH}. We note that while $\dot{E}_{+}$ exhibits no discernible scaling behavior, $\dot{E}_{-}$ exhibits power law scaling as $\epsilon \to 0$, which in Fig. \ref{fig:EFH} corresponds to $\sim \epsilon^{5/4}$. This behavior for $\dot{E}_{-}$ has been previously observed for near-extremal Kerr black holes, with a power scaling of $\sim \epsilon^{2/3}$ \cite{gralla_2015}. 
    
    In the same figures, we compared our numerical data with the slow-motion analytic formulas for the energy fluxes derived by \citeauthor{bini_2016} \cite{bini_2016}. While the qualitative behavior of \citeauthor{bini_2016}'s formula is similar to our numerical results for $\dot{E}_{-}$, that cannot be said for $\dot{E}_{+}$. The qualitative behavior exhibited by \citeauthor{bini_2016}'s formula for $\dot{E}_{+}$ is opposite to that of our numerical results, and the disagreement worsens as $\epsilon \to 0$. 
    
\begin{table*}[tb]
\caption{Energy flux towards infinity for various values of $\epsilon$ and $\Omega$.}
\label{tab:EFI}
\centering
\begin{tabular}[t]{| c | c c c c c c |}
\hline
$(M\Omega)^{-2/3}$ & $\epsilon = 1$ & $\epsilon = 0.5$ & $\epsilon = 0.3$ & $\epsilon = 0.1$ & $\epsilon = 0.001 M$ & $\epsilon = 0$ \\
\hline
$10$ & $3.1206577 \times 10^{-5}$ & $3.0536344 \times 10^{-5}$ & $2.9879739 \times 10^{-5}$ & $2.8982194 \times 10^{-5}$ & $2.8441983 \times 10^{-5}$ & $2.8436184 \times 10^{-5}$ \\
$20$ & $1.9825103 \times 10^{-6}$ & $1.9637653 \times 10^{-6}$ & $1.9455641 \times 10^{-6}$ & $1.9209679 \times 10^{-6}$ & $1.9063354 \times 10^{-6}$ & $1.9061791 \times 10^{-6}$ \\
$30$ & $3.9617935 \times 10^{-7}$ & $3.9378347 \times 10^{-7}$ & $3.9146543 \times 10^{-7}$ & $3.8834648 \times 10^{-7}$ & $3.8649863 \times 10^{-7}$ & $3.8647892 \times 10^{-7}$ \\
$50$ & $5.1966962 \times 10^{-8}$ & $5.1784602 \times 10^{-8}$ & $5.1608700 \times 10^{-8}$ & $5.1372867 \times 10^{-8}$ & $5.1233611 \times 10^{-8}$ & $5.1232128 \times 10^{-8}$ \\
$70$ & $1.3610616 \times 10^{-8}$ & $1.3576991 \times 10^{-8}$ & $1.3544600 \times 10^{-8}$ & $1.3501241 \times 10^{-8}$ & $1.3475676 \times 10^{-8}$ & $1.3475403 \times 10^{-8}$ \\
$100$ & $3.2846188 \times 10^{-9}$ & $3.2790005 \times 10^{-9}$ & $3.2735939 \times 10^{-9}$ & $3.2663651 \times 10^{-9}$ & $3.2621073 \times 10^{-9}$ & $3.2620620 \times 10^{-9}$ \\
\hline 
\end{tabular}
\end{table*}

\begin{table*}[tb]
\caption{Energy flux down the black hole for various values of $\epsilon$ and $\Omega$.}
\label{tab:EFH}
\centering
\begin{tabular}[t]{| c | c c c c c c |}
\hline
$(M\Omega)^{-2/3}$ & $\epsilon = 1$ & $\epsilon = 0.5$ & $\epsilon = 0.3$ & $\epsilon = 0.1$ & $\epsilon = 0.001$ & $\epsilon = 0$ \\
\hline
$10$ & $1.7007594 \times 10^{-7}$ & $1.1483202 \times 10^{-7}$ & $6.8214390 \times 10^{-8}$ & $1.3789999 \times 10^{-9}$ & $1.7818316 \times 10^{-10}$ & $5.9210431 \times 10^{-11}$\\
$20$ & $1.1596629 \times 10^{-9}$ & $7.6985329 \times 10^{-10}$ & $4.4908386 \times 10^{-10}$ & $1.2016755\times 10^{-10}$ & $7.3213635 \times 10^{-13}$ & $4.1611347 \times 10^{-14}$ \\
$30$ & $6.5341677 \times 10^{-11}$ & $4.3136903 \times 10^{-11}$ & $2.5022277 \times 10^{-11}$ & $6.6396635 \times 10^{-12}$ & $3.8178273 \times 10^{-14}$ & $6.5765960 \times 10^{-16}$ \\
$50$ & $1.7776656 \times 10^{-12}$ & $1.1683680 \times 10^{-12}$ & $6.7476326 \times 10^{-13}$ & $1.7794278 \times 10^{-13}$ & $1.0011889 \times 10^{-15}$ & $3.7180270 \times 10^{-18}$ \\
$70$ & $1.6665148 \times 10^{-13}$ & $1.0932322 \times 10^{-13}$ & $6.3019850 \times 10^{-14}$ & $1.6576708 \times 10^{-14}$ & $9.2830571 \times 10^{-17}$ & $1.2515752 \times 10^{-19}$ \\
$100$ & $1.3604719 \times 10^{-14}$ & $8.9119542 \times 10^{-15}$ & $5.1302207 \times 10^{-15}$ & $1.3469136 \times 10^{-15}$ & $7.5253293 \times 10^{-18}$ & $3.4688374 \times 10^{-21}$ \\
\hline 
\end{tabular}
\end{table*}

\subsection{Dissipative component of the self-force}

For circular orbits, the dissipative components of the scalar self-force are $F_t$, and $F_{\phi}$. We note that due to $u^{\alpha} F_{\alpha}=0$ in the circular orbit case, there is a simple relationship between the dissipative components of the self-force:
\begin{equation}
	F_t + \Omega F_{\phi} = 0. \label{eq:FtFphi}
\end{equation} 
This relationship indicates that we need only one component to calculate. In this work, we choose to calculate $F_t$.

For our set-up, the local energy dissipation must be accounted for by the energy fluxes towards infinity, and down the black hole. This energy balance relation can be expressed in terms of the self-force
\begin{equation}
	F_t = \mu u^t\dot{E}_{\mathrm{total}}. \label{eq:EBalance}
\end{equation}
This allows us to test our computation of $F_t$ by verifying that our numerical results satisfy Eq.~\eqref{eq:EBalance}. 

	Sample numerical results for $F_t$ are presented in Table \ref{tab:Ft}. As a check, we compared our Schwarzschild results $(\epsilon=1)$ with those of \citeauthor{warburton_2010} \cite{warburton_2010}, and we are in agreement to all significant figures presented. Looking at our results, we see that as the black hole approaches extremality $(\epsilon \to 0)$, the dissipative self-force decreases. One concludes from this that the black hole charge suppresses local energy dissipation.

\begin{figure}[tb]
\includegraphics[width=\linewidth]{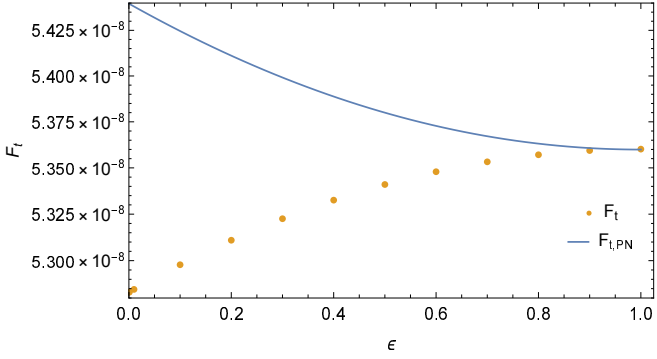}
\caption{Linear-log plot of $\epsilon$ vs $F_t$, for $(M\Omega)^{-2/3} = 50$. The orange dots represent the numerical results, while the solid blue line represents \citeauthor{bini_2016}'s slow-motion formula for the dissipative self-force. Disagreement between the two results increases significantly as $\epsilon \to 0$.}
\label{fig:Ft}
\end{figure}

	We also compared our results to the slow-motion formula for the dissipative self-force derived by \citeauthor{bini_2016} \cite{bini_2016}, presented in Fig. \ref{fig:Ft}. We see that the qualitative behavior of our results completely differs from \citeauthor{bini_2016}'s formula, which worsens as $\epsilon \to 0$. We note that this discrepancy in qualitative behavior is also present for $\dot{E}_{+}$, as presented in Fig. \ref{fig:EFI}.
    
\begin{table}[tb]
\caption{Comparison of the relative energy balance error for $(M\Omega)^{-2/3}=50$}
\label{tab:EBAL}
\begin{tabular}[t]{| c | c c |}
\hline
$\epsilon$ & Numerical Result & PN Result \\
\hline
$0$ & $4.1037989 \times 10^{-17}$ & $2.9851985 \times 10^{-2}$ \\
$0.001$ & $4.1160241 \times 10^{-17}$ & $2.9853170 \times 10^{-2}$ \\
$0.100$ & $4.4873798 \times 10^{-17}$ & $2.9965030 \times 10^{-2}$ \\
$0.300$ & $4.6552475 \times 10^{-17}$ & $3.0158060 \times 10^{-2}$ \\
$0.500$ & $4.6038684 \times 10^{-17}$ & $3.0305051 \times 10^{-2}$ \\
$0.800$ & $4.6637857 \times 10^{-17}$ & $3.0435251 \times 10^{-2}$ \\
$1.000$ & $4.6593530 \times 10^{-17}$ & $3.0460219 \times 10^{-2}$ \\
\hline
\end{tabular}
\end{table}
    
    As a consistency check, we then examined the energy balance relation exhibited by our numerical results and \citeauthor{bini_2016}'s slow-motion formulas. Sample data are presented in Table \ref{tab:EBAL}. While it is expected that energy balance will be better satisfied by numerical calculations compared to PN calculations, we note that the energy balance error exhibit by \citeauthor{bini_2016}'s formulas are of relative 3 PN order, one order higher than the expected error in their formulas.
    
\begin{table*}[tb]
\caption{Dissipative component of the self-force for various values of $\epsilon$ and $\Omega$.}
\label{tab:Ft}
\centering
\begin{tabular}[t]{| c | c c c c c c |}
\hline
$(M\Omega)^{-2/3}$ & $\epsilon = 1$ & $\epsilon = 0.5$ & $\epsilon = 0.3$ & $\epsilon = 0.1$ & $\epsilon = 0.001$ & $\epsilon = 0$ \\
\hline
$10$ & $3.7502273 \times 10^{-5}$ & $3.6569364 \times 10^{-5}$ & $3.5667917 \times 10^{-5}$ & $3.4458354 \times 10^{-5}$ & $3.3746536 \times 10^{-5}$ & $3.3739003 \times 10^{-5}$ \\
$20$ & $2.1515922 \times 10^{-6}$ & $2.1300513 \times 10^{-6}$ & $2.1092157 \times 10^{-6}$ & $2.0812004 \times 10^{-6}$ & $2.0646315 \times 10^{-6}$ & $2.0644552 \times 10^{-6}$ \\
$30$ & $4.1767858 \times 10^{-7}$ & $4.1506547 \times 10^{-7}$ & $4.1254172 \times 10^{-7}$ & $4.0915372 \times 10^{-7}$ & $4.0715177 \times 10^{-7}$ & $4.0712045 \times 10^{-7}$ \\
$50$ & $5.3601662 \times 10^{-8}$ & $5.3410098 \times 10^{-8}$ & $5.3225439 \times 10^{-8}$ & $5.2978070 \times 10^{-8}$ & $5.2832143 \times 10^{-8}$ & $5.2830590 \times 10^{-8}$ \\
$70$ & $1.3912164 \times 10^{-8}$ & $1.3877366 \times 10^{-8}$ & $1.3843857 \times 10^{-8}$ & $1.3799019 \times 10^{-8}$ & $1.3772594 \times 10^{-8}$ & $1.3772312 \times 10^{-8}$ \\
$100$ & $3.3350390 \times 10^{-9}$ & $3.3292867 \times 10^{-9}$ & $3.3237527 \times 10^{-9}$ & $3.1363538 \times 10^{-9}$ & $3.3119972 \times 10^{-9}$ & $3.3119508 \times 10^{-9}$ \\
\hline 
\end{tabular}
\end{table*}

\subsection{Conservative component of the self-force}

\begin{figure}[tb]
\includegraphics[width=\linewidth]{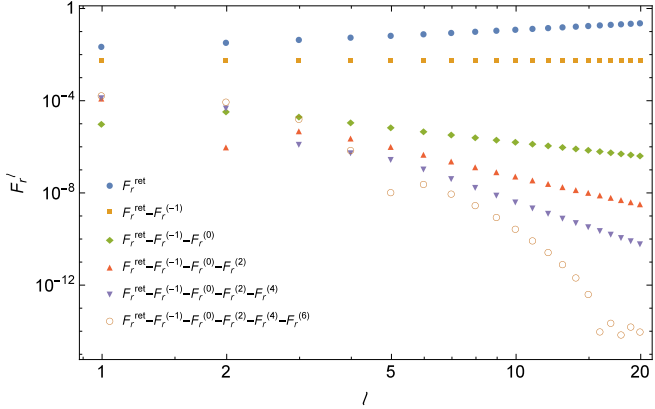}
\caption{$l$-Modes for the conservative self-force for $(M\Omega)^{-2/3}=10$ and $\epsilon=0.5$ , along with the results from the regularization by $A_r$, $B_r$, and additional regularization parameters ($D_r$, $E_r$, $F_r$) obtained from a numerical fit.}
\label{fig:Modesum}
\end{figure}

	For circular orbits, the conservative component of the self-force is contained entirely in $F_r$. The calculation of this conservative self-force is more complicated than the dissipative piece, as the mode-sum requires regularization. We then checked the effect of the regularization parameters on the mode-sum, as presented in Fig. \ref{fig:Modesum}. Looking at the high $l$-mode components, we see that the regularization parameters work as expected, leaving a residual field which exhibits $l^{-2}$ fall-off behavior.
    
    Sample numerical data for $F_r$ is presented in Table \ref{tab:Fr}. We compared our Schwarzschild results $(\epsilon = 1)$ with those of \citeauthor{diaz-rivera_2004} \cite{diaz-rivera_2004}, and we are in agreement up to six significant figures. Looking at our results, we see that as the black hole approaches extremality $(\epsilon \to 0)$, the conservative self-force decreases. This implies that the black hole charge suppresses the \textit{entire} self-force.
    
    We also compared our numerical results to the slow-motion formula for the conservative self-force derived by \citeauthor{bini_2016} \cite{bini_2016}. We present this in Fig. \ref{fig:Ft}, and we see that while \citeauthor{bini_2016}'s formula follow the same qualitative behavior of our results, we begin to deviate as $\epsilon \to 0$. 
    
\begin{table*}[tb]
\caption{Conservative component of the self-force for various values of $\epsilon$ and $\Omega$.}
\label{tab:Fr}
\centering
\begin{tabular}[t]{| c | c c c c c c |}
\hline
$(M\Omega)^{-2/3}$ & $\epsilon = 1$ & $\epsilon = 0.5$ & $\epsilon = 0.3$ & $\epsilon = 0.1$ & $\epsilon = 0.001$ & $\epsilon = 0$ \\
\hline
$10$ & $1.3784532 \times 10^{-5}$ & $1.3460616 \times 10^{-5}$ & $1.3154373 \times 10^{-5}$ & $1.2708095 \times 10^{-5}$ & $1.2434766 \times 10^{-5}$ & $1.2431961 \times 10^{-5}$ \\
$20$ & $4.9379089 \times 10^{-7}$ & $4.8179797 \times 10^{-7}$ & $4.7004549 \times 10^{-7}$ & $4.5399278 \times 10^{-7}$ & $4.4435692 \times 10^{-7}$ & $4.4425415 \times 10^{-7}$ \\
$30$ & $7.1719270 \times 10^{-8}$ & $7.0102613 \times 10^{-8}$ & $6.8530699 \times 10^{-8}$ & $6.6403455 \times 10^{-8}$ & $6.5136773 \times 10^{-8}$ & $6.5123254 \times 10^{-8}$ \\
$50$ & $6.3467922 \times 10^{-9}$ & $6.2203084 \times 10^{-9}$ & $6.0980083 \times 10^{-9}$ & $5.9335811 \times 10^{-9}$ & $5.8362462 \times 10^{-9}$ & $5.8352086 \times 10^{-9}$ \\
$70$ & $1.2845300 \times 10^{-9}$ & $1.2610382 \times 10^{-9}$ & $1.2383733 \times 10^{-9}$ & $1.2079790 \times 10^{-9}$ & $1.1900284 \times 10^{-9}$ & $1.1898372 \times 10^{-9}$ \\
$100$ & $2.3565036 \times 10^{-10}$ & $2.3171272 \times 10^{-10}$ & $2.2791964 \times 10^{-10}$ & $2.2284226 \times 10^{-10}$ & $2.1984858 \times 10^{-10}$ & $2.1981670 \times 10^{-10}$ \\
\hline 
\end{tabular}
\end{table*}

\begin{figure}[tb]
\includegraphics[width=\linewidth]{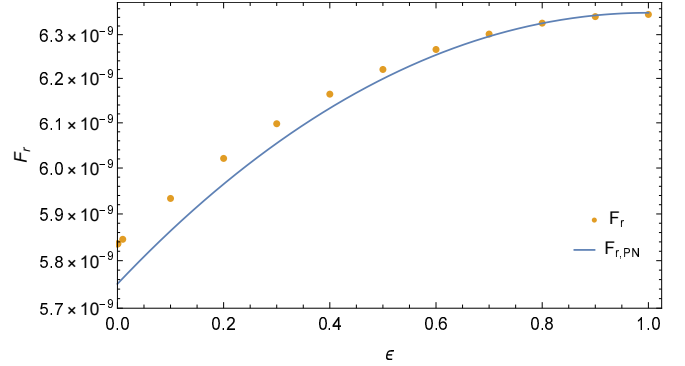}
\caption{Linear-log plot of $\epsilon$ vs $F_r$ for $(M\Omega)^{-2/3} = 50$. The orange dots represent the numerical results, while the solid blue line represents the results obtained from \citeauthor{bini_2016}'s slow-motion formula for the conservative self-force. Disagreement between the two results increases as $\epsilon \to 0$, however it is not as severe as compared to the dissipative self-force.}
\label{fig:Fr}
\end{figure}

\section{Conclusion}
In this work we presented the first mode-sum calculation of the self-force exerted on a particle in circular orbits about a Reissner-Nordstr\"{o}m black hole. We also present in this work regularization parameters $A_{\alpha}$, and $B_{\alpha}$ for circular orbits in Reissner-Nordstr\"{o}m spacetime.

We tested the validity of our results in various ways. The results for the Schwarzschild limit was found to agree with the results found in the literature \cite{warburton_2010,diaz-rivera_2004}. We confirmed numerically that the local energy dissipation is balanced out by the energy carried away by scalar waves towards infinity and down the event horizon. We also investigated the $l$-mode fall-off of the conservative self-force, and found that after subtracting the $A_r$ and $B_r$ regularization parameters the modes of the residual field fall off as $l^{-2}$, as expected.

Our results indicate that as the black hole's electric charge increases, the self-force decreases in magnitude. This dampening is notably drastic for the flux of scalar radiation towards the event horizon, where in near-extremal Reissner-Nordstr\"{o}m black holes, the scalar radiation flux scales as $\sim \epsilon^{5/4}$. This behavior is also seen for near-extremal Kerr black holes \cite{gralla_2015}, thus a more detailed calculation is recommended as a future study.

We also compared our results with the slow-motion formulas obtained by \citeauthor{bini_2016} \cite{bini_2016}. While our results agree in the Schwarzschild limit, they disagree for $Q \ne 0$, and as the electric charge increases, the disagreement between the results increases. We have yet to establish the reason for this disagreement.

We expect some of our results to have some bearing on future self-force studies in black hole solutions of scalar-tensor theories. The BBMB solution of conformal scalar-vacuum gravity is exactly the extremal Reissner-Nordstr\"{o}m geometry, so similar results might be obtained in situations where the scalar field in these alternative theories can be approximated as test fields. Other hairy black hole solutions have now been discovered in other scalar-tensor theories of gravity and some of these are also of Reissner-Nordstr\"{o}m form \cite{babichev_2017_hairybhs}. Self-force phenomenology in these theories remains completely uncharted, and so remains a promising area of future research. By exploring the scalar self-force from a minimally-coupled scalar field in the Reissner-Nordstr\"{o}m spacetime, we hope to have provided a useful guide and some benchmark numerical results for future self-force calculations in alternative theories of gravity.

\section{Acknowledgments}

This research is supported by the University of the Philippines OVPAA through Grant No.~OVPAA-BPhD-2016-13 and by the Department of Science and Technology Advanced Science and Technology Human Resources Development Program - National Science Consortium (DOST ASTHRDP-NSC).

\bibliography{references.bib}

\begin{thebibliography}{33}%
\makeatletter
\providecommand \@ifxundefined [1]{%
 \@ifx{#1\undefined}
}%
\providecommand \@ifnum [1]{%
 \ifnum #1\expandafter \@firstoftwo
 \else \expandafter \@secondoftwo
 \fi
}%
\providecommand \@ifx [1]{%
 \ifx #1\expandafter \@firstoftwo
 \else \expandafter \@secondoftwo
 \fi
}%
\providecommand \natexlab [1]{#1}%
\providecommand \enquote  [1]{``#1''}%
\providecommand \bibnamefont  [1]{#1}%
\providecommand \bibfnamefont [1]{#1}%
\providecommand \citenamefont [1]{#1}%
\providecommand \href@noop [0]{\@secondoftwo}%
\providecommand \href [0]{\begingroup \@sanitize@url \@href}%
\providecommand \@href[1]{\@@startlink{#1}\@@href}%
\providecommand \@@href[1]{\endgroup#1\@@endlink}%
\providecommand \@sanitize@url [0]{\catcode `\\12\catcode `\$12\catcode
  `\&12\catcode `\#12\catcode `\^12\catcode `\_12\catcode `\%12\relax}%
\providecommand \@@startlink[1]{}%
\providecommand \@@endlink[0]{}%
\providecommand \url  [0]{\begingroup\@sanitize@url \@url }%
\providecommand \@url [1]{\endgroup\@href {#1}{\urlprefix }}%
\providecommand \urlprefix  [0]{URL }%
\providecommand \Eprint [0]{\href }%
\providecommand \doibase [0]{http://dx.doi.org/}%
\providecommand \selectlanguage [0]{\@gobble}%
\providecommand \bibinfo  [0]{\@secondoftwo}%
\providecommand \bibfield  [0]{\@secondoftwo}%
\providecommand \translation [1]{[#1]}%
\providecommand \BibitemOpen [0]{}%
\providecommand \bibitemStop [0]{}%
\providecommand \bibitemNoStop [0]{.\EOS\space}%
\providecommand \EOS [0]{\spacefactor3000\relax}%
\providecommand \BibitemShut  [1]{\csname bibitem#1\endcsname}%
\let\auto@bib@innerbib\@empty
\bibitem [{\citenamefont {Amaro-Seoane}\ \emph {et~al.}(2012)\citenamefont
  {Amaro-Seoane} \emph {et~al.}}]{amaro-seoane_2012}%
  \BibitemOpen
  \bibfield  {author} {\bibinfo {author} {\bibfnamefont {P.}~\bibnamefont
  {Amaro-Seoane}} \emph {et~al.},\ }\href {\doibase
  10.1088/0264-9381/29/12/124016} {\bibfield  {journal} {\bibinfo  {journal}
  {Class. Quantum Grav.}\ }\textbf {\bibinfo {volume} {29}},\ \bibinfo {pages}
  {124016} (\bibinfo {year} {2012})}\BibitemShut {NoStop}%
\bibitem [{\citenamefont {Poisson}\ \emph {et~al.}(2011)\citenamefont
  {Poisson}, \citenamefont {Pound},\ and\ \citenamefont {Vega}}]{poisson_2011}%
  \BibitemOpen
  \bibfield  {author} {\bibinfo {author} {\bibfnamefont {E.}~\bibnamefont
  {Poisson}}, \bibinfo {author} {\bibfnamefont {A.}~\bibnamefont {Pound}}, \
  and\ \bibinfo {author} {\bibfnamefont {I.}~\bibnamefont {Vega}},\ }\href
  {\doibase 10.12942/lrr-2011-7} {\bibfield  {journal} {\bibinfo  {journal}
  {Living Rev. Relativity}\ }\textbf {\bibinfo {volume} {14}},\ \bibinfo
  {pages} {7} (\bibinfo {year} {2011})}\BibitemShut {NoStop}%
\bibitem [{\citenamefont {Barack}(2009)}]{barack_2009}%
  \BibitemOpen
  \bibfield  {author} {\bibinfo {author} {\bibfnamefont {L.}~\bibnamefont
  {Barack}},\ }\href {http://stacks.iop.org/0264-9381/26/i=21/a=213001}
  {\bibfield  {journal} {\bibinfo  {journal} {Class. Quantum Grav.}\ }\textbf
  {\bibinfo {volume} {26}},\ \bibinfo {pages} {213001} (\bibinfo {year}
  {2009})}\BibitemShut {NoStop}%
\bibitem [{\citenamefont {Wardell}(2015)}]{wardell_comp_2013}%
  \BibitemOpen
  \bibfield  {author} {\bibinfo {author} {\bibfnamefont {B.}~\bibnamefont
  {Wardell}},\ }\bibfield  {booktitle} {\emph {\bibinfo {booktitle}
  {{Proceedings, 524th WE-Heraeus-Seminar: Equations of Motion in Relativistic
  Gravity (EOM 2013): Bad Honnef, Germany, February 17-23, 2013}}},\ }\href
  {\doibase 10.1007/978-3-319-18335-0_14} {\bibfield  {journal} {\bibinfo
  {journal} {Fund. Theor. Phys.}\ }\textbf {\bibinfo {volume} {179}},\ \bibinfo
  {pages} {487} (\bibinfo {year} {2015})},\ \Eprint
  {http://arxiv.org/abs/1501.07322} {arXiv:1501.07322 [gr-qc]} \BibitemShut
  {NoStop}%
\bibitem [{\citenamefont {Drivas}\ and\ \citenamefont
  {Gralla}(2011)}]{drivas_2011}%
  \BibitemOpen
  \bibfield  {author} {\bibinfo {author} {\bibfnamefont {T.~D.}\ \bibnamefont
  {Drivas}}\ and\ \bibinfo {author} {\bibfnamefont {S.~E.}\ \bibnamefont
  {Gralla}},\ }\href {\doibase 10.1088/0264-9381/28/14/145025} {\bibfield
  {journal} {\bibinfo  {journal} {Class. Quantum Grav.}\ }\textbf {\bibinfo
  {volume} {28}},\ \bibinfo {pages} {145025} (\bibinfo {year}
  {2011})}\BibitemShut {NoStop}%
\bibitem [{\citenamefont {Taylor}(2013)}]{taylor_2013}%
  \BibitemOpen
  \bibfield  {author} {\bibinfo {author} {\bibfnamefont {P.}~\bibnamefont
  {Taylor}},\ }\href {\doibase 10.1103/PhysRevD.87.024046} {\bibfield
  {journal} {\bibinfo  {journal} {Phys. Rev. D}\ }\textbf {\bibinfo {volume}
  {87}},\ \bibinfo {pages} {024046} (\bibinfo {year} {2013})}\BibitemShut
  {NoStop}%
\bibitem [{\citenamefont {Kuchar}\ \emph {et~al.}(2013)\citenamefont {Kuchar},
  \citenamefont {Poisson},\ and\ \citenamefont {Vega}}]{kuchar_2013}%
  \BibitemOpen
  \bibfield  {author} {\bibinfo {author} {\bibfnamefont {J.}~\bibnamefont
  {Kuchar}}, \bibinfo {author} {\bibfnamefont {E.}~\bibnamefont {Poisson}}, \
  and\ \bibinfo {author} {\bibfnamefont {I.}~\bibnamefont {Vega}},\ }\href
  {\doibase 10.1088/0264-9381/30/23/235033} {\bibfield  {journal} {\bibinfo
  {journal} {Class. Quantum Grav.}\ }\textbf {\bibinfo {volume} {30}},\
  \bibinfo {pages} {235033} (\bibinfo {year} {2013})}\BibitemShut {NoStop}%
\bibitem [{\citenamefont {Taylor}\ and\ \citenamefont
  {Flanagan}(2015)}]{taylor_2015}%
  \BibitemOpen
  \bibfield  {author} {\bibinfo {author} {\bibfnamefont {P.}~\bibnamefont
  {Taylor}}\ and\ \bibinfo {author} {\bibfnamefont {{\'E}.~{\'E}.}\
  \bibnamefont {Flanagan}},\ }\href {\doibase 10.1103/PhysRevD.92.084032}
  {\bibfield  {journal} {\bibinfo  {journal} {Phys. Rev. D}\ }\textbf {\bibinfo
  {volume} {92}},\ \bibinfo {pages} {084032} (\bibinfo {year}
  {2015})}\BibitemShut {NoStop}%
\bibitem [{\citenamefont {Harte}\ \emph {et~al.}(2016)\citenamefont {Harte},
  \citenamefont {Flanagan},\ and\ \citenamefont {Taylor}}]{harte_2016}%
  \BibitemOpen
  \bibfield  {author} {\bibinfo {author} {\bibfnamefont {A.~I.}\ \bibnamefont
  {Harte}}, \bibinfo {author} {\bibfnamefont {{\'E}.~{\'E}.}\ \bibnamefont
  {Flanagan}}, \ and\ \bibinfo {author} {\bibfnamefont {P.}~\bibnamefont
  {Taylor}},\ }\href {\doibase 10.1103/PhysRevD.93.124054} {\bibfield
  {journal} {\bibinfo  {journal} {Phys. Rev. D}\ }\textbf {\bibinfo {volume}
  {93}},\ \bibinfo {pages} {124054} (\bibinfo {year} {2016})}\BibitemShut
  {NoStop}%
\bibitem [{\citenamefont {Harte}\ \emph {et~al.}(2017)\citenamefont {Harte},
  \citenamefont {Taylor},\ and\ \citenamefont {Flanagan}}]{harte_2017}%
  \BibitemOpen
  \bibfield  {author} {\bibinfo {author} {\bibfnamefont {A.~I.}\ \bibnamefont
  {Harte}}, \bibinfo {author} {\bibfnamefont {P.}~\bibnamefont {Taylor}}, \
  and\ \bibinfo {author} {\bibfnamefont {{\'{E}}.~{\'{E}}.}\ \bibnamefont
  {Flanagan}},\ }\href@noop {} {\  (\bibinfo {year} {2017})},\ \Eprint
  {http://arxiv.org/abs/1708.07813} {arXiv:1708.07813 [gr-qc]} \BibitemShut
  {NoStop}%
\bibitem [{\citenamefont {Zimmerman}(2015)}]{zimmerman_2015}%
  \BibitemOpen
  \bibfield  {author} {\bibinfo {author} {\bibfnamefont {P.}~\bibnamefont
  {Zimmerman}},\ }\href {\doibase 10.1103/PhysRevD.92.064051} {\bibfield
  {journal} {\bibinfo  {journal} {Phys. Rev. D}\ }\textbf {\bibinfo {volume}
  {92}},\ \bibinfo {pages} {064051} (\bibinfo {year} {2015})}\BibitemShut
  {NoStop}%
\bibitem [{\citenamefont {Torii}\ \emph {et~al.}(2001)\citenamefont {Torii},
  \citenamefont {Maeda},\ and\ \citenamefont {Narita}}]{torii_2001}%
  \BibitemOpen
  \bibfield  {author} {\bibinfo {author} {\bibfnamefont {T.}~\bibnamefont
  {Torii}}, \bibinfo {author} {\bibfnamefont {K.}~\bibnamefont {Maeda}}, \ and\
  \bibinfo {author} {\bibfnamefont {M.}~\bibnamefont {Narita}},\ }\href
  {\doibase 10.1103/PhysRevD.64.044007} {\bibfield  {journal} {\bibinfo
  {journal} {Phys. Rev. D}\ }\textbf {\bibinfo {volume} {64}},\ \bibinfo
  {pages} {044007} (\bibinfo {year} {2001})}\BibitemShut {NoStop}%
\bibitem [{\citenamefont {Kanti}\ \emph {et~al.}(1996)\citenamefont {Kanti},
  \citenamefont {Mavromatos}, \citenamefont {Rizos}, \citenamefont {Tamvakis},\
  and\ \citenamefont {Winstanley}}]{kanti_1996}%
  \BibitemOpen
  \bibfield  {author} {\bibinfo {author} {\bibfnamefont {P.}~\bibnamefont
  {Kanti}}, \bibinfo {author} {\bibfnamefont {N.~E.}\ \bibnamefont
  {Mavromatos}}, \bibinfo {author} {\bibfnamefont {J.}~\bibnamefont {Rizos}},
  \bibinfo {author} {\bibfnamefont {K.}~\bibnamefont {Tamvakis}}, \ and\
  \bibinfo {author} {\bibfnamefont {E.}~\bibnamefont {Winstanley}},\ }\href
  {\doibase 10.1103/PhysRevD.54.5049} {\bibfield  {journal} {\bibinfo
  {journal} {Phys. Rev. D}\ }\textbf {\bibinfo {volume} {54}},\ \bibinfo
  {pages} {5049} (\bibinfo {year} {1996})}\BibitemShut {NoStop}%
\bibitem [{\citenamefont {Sotiriou}\ and\ \citenamefont
  {Zhou}(2014)}]{sotiriou_2014}%
  \BibitemOpen
  \bibfield  {author} {\bibinfo {author} {\bibfnamefont {T.~P.}\ \bibnamefont
  {Sotiriou}}\ and\ \bibinfo {author} {\bibfnamefont {S.-Y.}\ \bibnamefont
  {Zhou}},\ }\href {\doibase 10.1103/PhysRevD.90.124063} {\bibfield  {journal}
  {\bibinfo  {journal} {Phys. Rev. D}\ }\textbf {\bibinfo {volume} {90}},\
  \bibinfo {pages} {124063} (\bibinfo {year} {2014})}\BibitemShut {NoStop}%
\bibitem [{\citenamefont {Antoniou}\ \emph {et~al.}(2018)\citenamefont
  {Antoniou}, \citenamefont {Bakopoulos},\ and\ \citenamefont
  {Kanti}}]{antoniou_2018}%
  \BibitemOpen
  \bibfield  {author} {\bibinfo {author} {\bibfnamefont {G.}~\bibnamefont
  {Antoniou}}, \bibinfo {author} {\bibfnamefont {A.}~\bibnamefont
  {Bakopoulos}}, \ and\ \bibinfo {author} {\bibfnamefont {P.}~\bibnamefont
  {Kanti}},\ }\href {\doibase 10.1103/PhysRevLett.120.131102} {\bibfield
  {journal} {\bibinfo  {journal} {Phys. Rev. Lett.}\ }\textbf {\bibinfo
  {volume} {120}},\ \bibinfo {pages} {131102} (\bibinfo {year}
  {2018})}\BibitemShut {NoStop}%
\bibitem [{\citenamefont {Antoniou}\ \emph {et~al.}(2017)\citenamefont
  {Antoniou}, \citenamefont {Bakopoulos},\ and\ \citenamefont
  {Kanti}}]{antoniou_2017}%
  \BibitemOpen
  \bibfield  {author} {\bibinfo {author} {\bibfnamefont {G.}~\bibnamefont
  {Antoniou}}, \bibinfo {author} {\bibfnamefont {A.}~\bibnamefont
  {Bakopoulos}}, \ and\ \bibinfo {author} {\bibfnamefont {P.}~\bibnamefont
  {Kanti}},\ }\href@noop {} {\  (\bibinfo {year} {2017})},\ \Eprint
  {http://arxiv.org/abs/1711.07431} {arXiv:1711.07431 [hep-th]} \BibitemShut
  {NoStop}%
\bibitem [{\citenamefont {Babichev}\ \emph {et~al.}(2017)\citenamefont
  {Babichev}, \citenamefont {Charmousis},\ and\ \citenamefont
  {Lehébel}}]{babichev_2017_hairybhs}%
  \BibitemOpen
  \bibfield  {author} {\bibinfo {author} {\bibfnamefont {E.}~\bibnamefont
  {Babichev}}, \bibinfo {author} {\bibfnamefont {C.}~\bibnamefont
  {Charmousis}}, \ and\ \bibinfo {author} {\bibfnamefont {A.}~\bibnamefont
  {Lehébel}},\ }\href {\doibase 10.1088/1475-7516/2017/04/027} {\bibfield
  {journal} {\bibinfo  {journal} {JCAP}\ }\textbf {\bibinfo {volume} {1704}},\
  \bibinfo {pages} {027} (\bibinfo {year} {2017})},\ \Eprint
  {http://arxiv.org/abs/1702.01938} {arXiv:1702.01938 [gr-qc]} \BibitemShut
  {NoStop}%
\bibitem [{\citenamefont {Bocharova}\ \emph {et~al.}(1970)\citenamefont
  {Bocharova}, \citenamefont {Bronnikov},\ and\ \citenamefont
  {Melnikov}}]{bocharova_1970}%
  \BibitemOpen
  \bibfield  {author} {\bibinfo {author} {\bibfnamefont {N.~M.}\ \bibnamefont
  {Bocharova}}, \bibinfo {author} {\bibfnamefont {K.~A.}\ \bibnamefont
  {Bronnikov}}, \ and\ \bibinfo {author} {\bibfnamefont {V.~N.}\ \bibnamefont
  {Melnikov}},\ }\href@noop {} {\bibfield  {journal} {\bibinfo  {journal}
  {Vestn. Mosk. Univ. Fiz. Astron.}\ }\textbf {\bibinfo {volume} {6}},\
  \bibinfo {pages} {706} (\bibinfo {year} {1970})}\BibitemShut {NoStop}%
\bibitem [{\citenamefont {Bekenstein}(1974)}]{bekenstein_1974}%
  \BibitemOpen
  \bibfield  {author} {\bibinfo {author} {\bibfnamefont {J.~D.}\ \bibnamefont
  {Bekenstein}},\ }\href@noop {} {\bibfield  {journal} {\bibinfo  {journal}
  {Ann. Phys.}\ }\textbf {\bibinfo {volume} {82}},\ \bibinfo {pages} {535}
  (\bibinfo {year} {1974})}\BibitemShut {NoStop}%
\bibitem [{\citenamefont {Bekenstein}(1975)}]{bekenstein_1975}%
  \BibitemOpen
  \bibfield  {author} {\bibinfo {author} {\bibfnamefont {J.~D.}\ \bibnamefont
  {Bekenstein}},\ }\href@noop {} {\bibfield  {journal} {\bibinfo  {journal}
  {Ann. Phys.}\ }\textbf {\bibinfo {volume} {91}},\ \bibinfo {pages} {75}
  (\bibinfo {year} {1975})}\BibitemShut {NoStop}%
\bibitem [{\citenamefont {Bini}\ \emph {et~al.}(2016)\citenamefont {Bini},
  \citenamefont {Carvalho},\ and\ \citenamefont {Geralico}}]{bini_2016}%
  \BibitemOpen
  \bibfield  {author} {\bibinfo {author} {\bibfnamefont {D.}~\bibnamefont
  {Bini}}, \bibinfo {author} {\bibfnamefont {G.~G.}\ \bibnamefont {Carvalho}},
  \ and\ \bibinfo {author} {\bibfnamefont {A.}~\bibnamefont {Geralico}},\
  }\href@noop {} {\bibfield  {journal} {\bibinfo  {journal} {Phys. Rev. D}\
  }\textbf {\bibinfo {volume} {94}},\ \bibinfo {pages} {124028} (\bibinfo
  {year} {2016})}\BibitemShut {NoStop}%
\bibitem [{\citenamefont {Barack}\ and\ \citenamefont
  {Ori}(2000)}]{barack_2000}%
  \BibitemOpen
  \bibfield  {author} {\bibinfo {author} {\bibfnamefont {L.}~\bibnamefont
  {Barack}}\ and\ \bibinfo {author} {\bibfnamefont {A.}~\bibnamefont {Ori}},\
  }\href@noop {} {\bibfield  {journal} {\bibinfo  {journal} {Phys. Rev. D}\
  }\textbf {\bibinfo {volume} {61}},\ \bibinfo {pages} {061502} (\bibinfo
  {year} {2000})}\BibitemShut {NoStop}%
\bibitem [{\citenamefont {Burko}(2000)}]{burko_2000_b}%
  \BibitemOpen
  \bibfield  {author} {\bibinfo {author} {\bibfnamefont {L.~M.}\ \bibnamefont
  {Burko}},\ }\href {\doibase 10.1103/PhysRevLett.84.4529} {\bibfield
  {journal} {\bibinfo  {journal} {Phys. Rev. Lett.}\ }\textbf {\bibinfo
  {volume} {84}},\ \bibinfo {pages} {4529} (\bibinfo {year}
  {2000})}\BibitemShut {NoStop}%
\bibitem [{\citenamefont {Barack}\ and\ \citenamefont
  {Ori}(2003)}]{barack_2003}%
  \BibitemOpen
  \bibfield  {author} {\bibinfo {author} {\bibfnamefont {L.}~\bibnamefont
  {Barack}}\ and\ \bibinfo {author} {\bibfnamefont {A.}~\bibnamefont {Ori}},\
  }\href {\doibase 10.1103/PhysRevLett.90.111101} {\bibfield  {journal}
  {\bibinfo  {journal} {Phys. Rev. Lett.}\ }\textbf {\bibinfo {volume} {90}},\
  \bibinfo {pages} {111101} (\bibinfo {year} {2003})}\BibitemShut {NoStop}%
\bibitem [{\citenamefont {Barack}\ \emph {et~al.}(2002)\citenamefont {Barack},
  \citenamefont {Mino}, \citenamefont {Nakano}, \citenamefont {Ori},\ and\
  \citenamefont {Sasaki}}]{barack_2002}%
  \BibitemOpen
  \bibfield  {author} {\bibinfo {author} {\bibfnamefont {L.}~\bibnamefont
  {Barack}}, \bibinfo {author} {\bibfnamefont {Y.}~\bibnamefont {Mino}},
  \bibinfo {author} {\bibfnamefont {H.}~\bibnamefont {Nakano}}, \bibinfo
  {author} {\bibfnamefont {A.}~\bibnamefont {Ori}}, \ and\ \bibinfo {author}
  {\bibfnamefont {M.}~\bibnamefont {Sasaki}},\ }\href {\doibase
  10.1103/PhysRevLett.88.091101} {\bibfield  {journal} {\bibinfo  {journal}
  {Phys. Rev. Lett.}\ }\textbf {\bibinfo {volume} {88}},\ \bibinfo {pages}
  {091101} (\bibinfo {year} {2002})}\BibitemShut {NoStop}%
\bibitem [{\citenamefont {Haas}\ and\ \citenamefont
  {Poisson}(2006)}]{Haas:2006ne}%
  \BibitemOpen
  \bibfield  {author} {\bibinfo {author} {\bibfnamefont {R.}~\bibnamefont
  {Haas}}\ and\ \bibinfo {author} {\bibfnamefont {E.}~\bibnamefont {Poisson}},\
  }\href {\doibase 10.1103/PhysRevD.74.044009} {\bibfield  {journal} {\bibinfo
  {journal} {Phys. Rev. D}\ }\textbf {\bibinfo {volume} {74}},\ \bibinfo
  {pages} {044009} (\bibinfo {year} {2006})},\ \Eprint
  {http://arxiv.org/abs/gr-qc/0605077} {arXiv:gr-qc/0605077 [gr-qc]}
  \BibitemShut {NoStop}%
\bibitem [{\citenamefont {Heffernan}\ \emph {et~al.}(2012)\citenamefont
  {Heffernan}, \citenamefont {Ottewill},\ and\ \citenamefont
  {Wardell}}]{Heffernan:2012su}%
  \BibitemOpen
  \bibfield  {author} {\bibinfo {author} {\bibfnamefont {A.}~\bibnamefont
  {Heffernan}}, \bibinfo {author} {\bibfnamefont {A.}~\bibnamefont {Ottewill}},
  \ and\ \bibinfo {author} {\bibfnamefont {B.}~\bibnamefont {Wardell}},\ }\href
  {\doibase 10.1103/PhysRevD.86.104023} {\bibfield  {journal} {\bibinfo
  {journal} {Phys. Rev. D}\ }\textbf {\bibinfo {volume} {86}},\ \bibinfo
  {pages} {104023} (\bibinfo {year} {2012})},\ \Eprint
  {http://arxiv.org/abs/1204.0794} {arXiv:1204.0794 [gr-qc]} \BibitemShut
  {NoStop}%
\bibitem [{\citenamefont {Heffernan}\ \emph {et~al.}(2014)\citenamefont
  {Heffernan}, \citenamefont {Ottewill},\ and\ \citenamefont
  {Wardell}}]{Heffernan:2012vj}%
  \BibitemOpen
  \bibfield  {author} {\bibinfo {author} {\bibfnamefont {A.}~\bibnamefont
  {Heffernan}}, \bibinfo {author} {\bibfnamefont {A.}~\bibnamefont {Ottewill}},
  \ and\ \bibinfo {author} {\bibfnamefont {B.}~\bibnamefont {Wardell}},\ }\href
  {\doibase 10.1103/PhysRevD.89.024030} {\bibfield  {journal} {\bibinfo
  {journal} {Phys. Rev. D}\ }\textbf {\bibinfo {volume} {89}},\ \bibinfo
  {pages} {024030} (\bibinfo {year} {2014})},\ \Eprint
  {http://arxiv.org/abs/1211.6446} {arXiv:1211.6446 [gr-qc]} \BibitemShut
  {NoStop}%
\bibitem [{\citenamefont {Heffernan}(2012)}]{Heffernan2014}%
  \BibitemOpen
  \bibfield  {author} {\bibinfo {author} {\bibfnamefont {A.}~\bibnamefont
  {Heffernan}},\ }\emph {\bibinfo {title} {{The Self-Force Problem: Local
  Behavior of the Detweiler-Whiting Singular Field}}},\ \href
  {https://inspirehep.net/record/1287046/files/arXiv:1403.6177.pdf} {Ph.D.
  thesis},\ \bibinfo  {school} {University Coll. Dublin} (\bibinfo {year}
  {2012}),\ \Eprint {http://arxiv.org/abs/1403.6177} {arXiv:1403.6177 [gr-qc]}
  \BibitemShut {NoStop}%
\bibitem [{\citenamefont {Detweiler}\ and\ \citenamefont
  {Whiting}(2003)}]{detweiler_2003}%
  \BibitemOpen
  \bibfield  {author} {\bibinfo {author} {\bibfnamefont {S.}~\bibnamefont
  {Detweiler}}\ and\ \bibinfo {author} {\bibfnamefont {B.~F.}\ \bibnamefont
  {Whiting}},\ }\href {\doibase 10.1103/PhysRevD.67.024025} {\bibfield
  {journal} {\bibinfo  {journal} {Phys. Rev. D}\ }\textbf {\bibinfo {volume}
  {67}},\ \bibinfo {pages} {024025} (\bibinfo {year} {2003})}\BibitemShut
  {NoStop}%
\bibitem [{\citenamefont {Gralla}\ \emph {et~al.}(2015)\citenamefont {Gralla},
  \citenamefont {Porfyriadis},\ and\ \citenamefont {Warburton}}]{gralla_2015}%
  \BibitemOpen
  \bibfield  {author} {\bibinfo {author} {\bibfnamefont {S.~E.}\ \bibnamefont
  {Gralla}}, \bibinfo {author} {\bibfnamefont {A.~P.}\ \bibnamefont
  {Porfyriadis}}, \ and\ \bibinfo {author} {\bibfnamefont {N.}~\bibnamefont
  {Warburton}},\ }\href {\doibase 10.1103/PhysRevD.92.064029} {\bibfield
  {journal} {\bibinfo  {journal} {Phys. Rev. D}\ }\textbf {\bibinfo {volume}
  {92}},\ \bibinfo {pages} {064029} (\bibinfo {year} {2015})}\BibitemShut
  {NoStop}%
\bibitem [{\citenamefont {Warburton}\ and\ \citenamefont
  {Barack}(2010)}]{warburton_2010}%
  \BibitemOpen
  \bibfield  {author} {\bibinfo {author} {\bibfnamefont {N.}~\bibnamefont
  {Warburton}}\ and\ \bibinfo {author} {\bibfnamefont {L.}~\bibnamefont
  {Barack}},\ }\href {\doibase 10.1103/PhysRevD.81.084039} {\bibfield
  {journal} {\bibinfo  {journal} {Phys. Rev. D}\ }\textbf {\bibinfo {volume}
  {81}},\ \bibinfo {pages} {084039} (\bibinfo {year} {2010})}\BibitemShut
  {NoStop}%
\bibitem [{\citenamefont {Diaz-Rivera}\ \emph {et~al.}(2004)\citenamefont
  {Diaz-Rivera}, \citenamefont {Messaritaki}, \citenamefont {Whiting},\ and\
  \citenamefont {Detweiler}}]{diaz-rivera_2004}%
  \BibitemOpen
  \bibfield  {author} {\bibinfo {author} {\bibfnamefont {L.~M.}\ \bibnamefont
  {Diaz-Rivera}}, \bibinfo {author} {\bibfnamefont {E.}~\bibnamefont
  {Messaritaki}}, \bibinfo {author} {\bibfnamefont {B.~F.}\ \bibnamefont
  {Whiting}}, \ and\ \bibinfo {author} {\bibfnamefont {S.}~\bibnamefont
  {Detweiler}},\ }\href@noop {} {\bibfield  {journal} {\bibinfo  {journal}
  {Phys. Rev. D}\ }\textbf {\bibinfo {volume} {70}},\ \bibinfo {pages} {124018}
  (\bibinfo {year} {2004})}\BibitemShut {NoStop}%
\end{thebibliography}%

\end{document}